# Ferromagnetism below 10 K in Mn doped BiTe

J.W.G. Bos[a], M. Lee[b], E. Morosan[a], H.W. Zandbergen[c], W.L. Lee[b], N.P. Ong[b], and R.J. Cava[a]

[a] Department of Chemistry, Princeton University, Princeton, New Jersey 08544, USA

[b] Department of Physics, Princeton University, Princeton, New Jersey 08544, USA

[c] National Centre for High Resolution Electron Microscopy, Delft Institute of Technology, Delft, The Netherlands.

Ferromagnetism is observed below 10 K in $[Bi_{0.75}Te_{0.125}Mn_{0.125}]Te$. This material has the BiTe structure, which is made from the stacking of two Te-Bi-Te-Bi-Te blocks and one Bi-Bi block per unit cell. Crystal structure analysis shows that Mn is localized in the $Bi_2$ blocks, and is accompanied by an equal amount of $Te_{Bi}$ anti-site occupancy in the $Bi_2Te_3$ blocks. These $Te_{Bi}$ anti-site defects greatly enhance the Mn solubility. This is demonstrated by comparison of the $[Bi_{1-x}Mn_x]Te$ and $[Bi_{1-2x}Te_xMn_x]Te$ series; in the former, the solubility is limited to $x = 0.067$, while the latter has $x_{max} = 0.125$. The magnetism in $[Bi_{1-x}Mn_x]Te$ changes little with x, while that for $[Bi_{1-2x}Te_xMn_x]Te$ shows a clear variation, leading to ferromagnetism for $x > 0.067$. Magnetic hysteresis and the anomalous Hall Effect are observed for the ferromagnetic samples.

**Introduction**

The integration of ferromagnetism into semiconductors is of great current interest.[1] Perhaps the best understood examples are molecular beam epitaxy (MBE) grown thin films of $Ga_{1-x}Mn_xAs$ and $In_{1-x}Mn_xAs$, where a few percent Mn substitution induces ferromagnetism with maximum Curie temperatures around 170 K.[2, 3] In this case, Mn acts as an acceptor, providing holes that mediate a ferromagnetic interaction between the local moments of the open $d$ shells in the Mn atoms. Carrier induced ferromagnetism has also been reported for $Pb_{1-x-y}Sn_yMn_xTe$[4], MBE grown p-doped $Cd_{1-x}Mn_xTe$ quantum wells[5], and $Zn_{1-x}Mn_xTe$ epilayers,[6] which have much lower Curie temperatures (below 5 K). Bulk samples of $Cd_{1-x}Mn_xTe$ and $Zn_{1-x}Mn_xTe$ typically show spin-glass behavior, but do show strong coupling between the sp bands and localized d electrons, evidenced, for example, by the giant spin-splitting of the bands or the magnetic field induced metal-insulator transition.[7] This does not lead to ferromagnetism, however, which only occurs after hole doping. In metallic systems, such as $Cu_{1-x}Mn_x$, the random substitution of Mn in the Cu host lattice leads to spin-glass behavior.[8]

Here we report the synthesis and characterization of Mn doped BiTe; ferromagnetism rather than spin-glass behavior is observed. Mn and Fe doped $Bi_2Te_3$ have been reported as have V, Cr, and Mn doped $Sb_2Te_3$. For single crystals of $Sb_{2-x}TM_xTe_3$ ferromagnetism is reported for TM = V ($T_c$ = 22 K, x = 0.03)[9] and Cr ($T_c$ = 20 K, x = 0.06),[10] while TM = Mn ($x_{max}$ = 0.045)[11] remains paramagnetic down to 2 K. For $Bi_{2-x}TM_xTe_3$, Curie temperatures of 12 K (TM = Fe, x = 0.08)[12] and 10 K (TM = Mn, x = 0.02 and 0.04)[13, 14] are reported. However, questions remain regarding the origin of the FM state in the latter two materials. For example, $Bi_{2-x}Fe_xTe_3$ has a maximum magnetization of only ~0.025 $\mu_B$/Fe (x = 0.08), and the transition temperatures for the two reported $Bi_{2-x}Mn_xTe_3$ compositions are identical. The studies so far have reported the magnetic and transport properties of these systems but contain no structural analysis. Here, we report the first case in this family of compounds where structural analysis proves that the Mn is incorporated within the semiconductor, and allows for its position within the lattice to be determined. The system reported is Mn-doped BiTe, which is part of the same chemical family as $Bi_2Te_3$.

BiTe belongs to the $(Bi_2Te_3)_m \cdot (Bi_2)_n$ homologous series, which is composed of different ratios of stacking of five-layer Te-Bi-Te-Bi-Te (5) blocks and two-layer Bi-Bi (2) blocks.[15] For example, $Bi_2Te_3$ has three rhombohedrally stacked (5) blocks per unit cell, and BiTe is built up from two (5) and one (2) blocks. Van der Waals gaps exist only between adjacent (5) blocks, making $Bi_2Te_3$ and BiTe true 2D materials. Band structure calculations on the isostructural and isoelectronic $(Bi_2Se_3)_m \cdot (Bi_2)_n$ homologous series indicate that $Bi_2Se_3$ is a narrow gap semiconductor, whereas Bi-Bi block containing materials are semimetals.[16,17] These calculations also confirm the absence of anionic Se-Se bonding in BiSe, and suggest the assignments of formal oxidation states of $(Bi^0)_2$ and $(Bi^{3+})_2(Se^{2-})_3$. The addition of the zero-valent $Bi_2$ blocks therefore does not change the charge balance in the $Bi_2Se_3$ layers and explains the formation of the $(Bi_2Se_3)_m \cdot (Bi_2)_n$ homologous series.

In the current work, structural analysis of Mn doped BiTe shows that Mn preferentially occupies sites in the $Bi_2$ blocks. The maximum Mn solubility in $[Bi_{1-x}Mn_x]Te$ is 0.067, which corresponds to a (2) block composition of $Bi_{0.8}Mn_{0.2}$. The $[Bi_{1-x}Mn_x]Te$ materials exhibit Curie-Weiss paramagnetism down to 5 K. Introduction of corresponding $Te_{Bi}$ anti-site occupancy in the (5) layer increases the solubility of Mn dramatically, and for $[Bi_{1-2x}Te_xMn_x]Te$ the maximum solubility is 0.125. Ferromagnetism is observed below 5 K for x = 0.100 and below 10 K for x = 0.125, corresponding to (2) block compositions of $Bi_{0.7}Mn_{0.3}$ and $Bi_{0.625}Mn_{0.375}$, respectively.

**Experimental**

Initial attempts to synthesize Mn intercalated $Bi_2Te_3$ with composition $Mn_{0.33}Bi_2Te_3$ resulted in the formation of a BiTe type phase. Detailed structural analysis (below) showed

Mn and an equal amount of Te to occupy different Bi sites. To confirm this, a series of compounds with formulas [Bi$_{1-2x}$Te$_x$Mn$_x$]Te and [Bi$_{1-x}$Mn$_x$]Te (0.033 ≤ x ≤ 0.167) was synthesized. The initial composition (corresponding to x = 0.125 in [Bi$_{1-2x}$Te$_x$Mn$_x$]Te) was obtained by heating small pieces of elemental Mn (99.95%), Bi (99.99%) and Te (99.99%) in the ratio 1:6:9 at 800 $^0$C for one day, followed by a two week anneal (with two intermediate regrindings) of homogenized pressed pellets at 575 $^0$C. The [Bi$_{1-2x}$Te$_x$Mn$_x$]Te and [Bi$_{1-x}$Mn$_x$]Te series were synthesized by heating intimately mixed Mn, Bi and Te powders slowly to 525 $^0$C, followed by a two week anneal of pressed pellets at 525 $^0$C (with two intermediate regrindings). All syntheses were done in vacuum sealed quartz tubes. Phase purity was checked by powder X-Ray diffraction (PXD) on a Bruker D8 Focus diffractometer with Cu K$_\alpha$ radiation and a diffracted beam monochromator. The GSAS suite of programs was used for Rietveld fitting of the PXD data.[18] A Pseudo-Voigt function was used to describe the peak shape. Elemental mapping using Energy Dispersive X-ray (EDX) analysis was performed on the samples. This mapping indicated that samples that were found to be single phase by X-ray diffraction displayed a homogeneous distribution of Mn, Bi and Te, and no detectable impurity phases to an estimated sensitivity of 1% (three times the noise in the background of the X-ray patterns). Elemental analysis using inductively coupled plasma atomic emission spectroscopy (ICP-AES) confirmed that the phases produced had the compositions expected from the starting compositions.

For [Bi$_{0.75}$Te$_{0.125}$Mn$_{0.125}$]Te, the temperature dependencies of the zero field cooled (ZFC) and field cooled (FC) magnetization were measured on a Quantum Design Magnetic Property Measurement System (MPMS) in an applied field of 100 Oe. The field dependence of the magnetization was measured on a Quantum Design Physical Property Measurement System (PPMS) fitted with an ACMS insert. The 2K M(H) hysteresis loop was measured on the MPMS. The temperature dependence of the sample resistivity was measured using a standard four point method. For $[Bi_{1-2x}Te_xMn_x]Te$ and $[Bi_{1-x}Mn_x]Te$ the temperature dependencies of the ZFC magnetic susceptibilities were measured on the PPMS in an applied field of 10 kOe. M(H) curves at 5 K were measured on the same instrument. Low temperature ZFC and FC susceptibilities (2-30 K, H = 100 Oe), and M(H) hysteresis curves at 2 K were collected on the MPMS.

**Crystal Structure**

Profile analysis of the X-ray powder diffraction pattern (Fig. 1) showed the single phase material of composition $Mn_{0.33}Bi_2Te_3$ to have the BiTe crystal structure (Fig. 2) with significantly smaller unit cell constants than are found for pure BiTe (Table 1). The crystal structure analysis showed that all the Mn and some of the Te occupy the Bi sites in BiTe. The formula of the compound is therefore best represented as $[Bi_{0.75}Te_{0.125}Mn_{0.125}]Te$. The structural model with Mn substituted in the $Bi_2$ block and $Te_{Bi}$ anti-site defects in the $Bi_2Te_3$ blocks resulted in the best Rietveld fits to the diffraction data (Table 1, $\chi^2 = 2.1$). Models with Mn (and $Te_{Bi}$ anti-sites) in both the $Bi_2$ and $Bi_2Te_3$ blocks were found to have significantly worse goodness of fit to the diffraction data. For example, a model with equal amounts of Mn and $Te_{Bi}$ on all Bi positions has $\chi^2 = 2.8$. Structure refinements in which Mn was omitted completely from the compound yielded $\chi^2 = 2.9$, showing the sensitivity of the diffraction data to the presence of the Mn. Table 1 presents the structural parameters obtained, and the agreement factors. These and the comparison of observed and calculated intensities shown in figure 1 indicate the high quality of the fit. The composition obtained by free refinement of the diffraction data is $[Bi_{0.75}Te_{0.122(3)}Mn_{0.128(3)}]Te$, in excellent agreement with the starting composition. This indicates that if any Mn is present in interstitial positions or secondary phases the amount would have to be minimal. The Mn atoms are preferentially located in the $Bi_2$ block, which has stoichiometry $(Bi_{0.62}Mn_{0.38})_2$. This high proportion of $Bi_2$ block occupancy explains why the earlier studies on TM doped $Sb_2Te_3$ and $Bi_2Te_3$ found that only a few atomic percent transition metal substitution is possible[9-14]: there are no $Sb_2$ or $Bi_2$ blocks in those compounds. The amount of Mn on site 1 is exactly balanced by the amount of $Te_{Bi}$ anti-sites on site 2 (see table 1). Interestingly, the $Te_{Bi}$ anti-site defects in the $Bi_2Te_3$ layers are all located on the Bi position closest to the $Bi_2$ layer. Deviations from 1:1 stoichiometry in BiTe, likely due to the presence of anti-site defects of the type reported here, have been observed previously in mineral samples, where $[Bi_{0.58}Te_{0.42}]Te$ and $Bi[Te_{0.75}Bi_{0.25}]$ have been reported with the BiTe structure.[19] The bond distances in $[Bi_{0.75}Te_{0.125}Mn_{0.125}]Te$ are changed considerably from those found in pure BiTe (Table 2), signaling a complex charge re-distribution upon Mn doping. The most obvious changes are the reduction of the Van der Waals gap between Te3-Te3 layers and the elongation of the (Bi/Te)1-Te2 bonds.

Diffraction analysis of the $[Bi_{1-2x}Te_xMn_x]Te$ series of samples, synthesized to conform to the Mn substitution and anti-site defect structure found in the detailed structural analysis of $[Bi_{0.75}Te_{0.125}Mn_{0.125}]Te$ showed $0.033 \leq x \leq 0.125$ to be single phase and to have the BiTe structure. The PXD patterns are given in Fig. 3. A systematic reduction in lattice constants with increasing x is found, confirming the presence of Mn substitution in a solid solution. For $x = 0.133$ and $0.167$, a $MnTe_2$ impurity is found, therefore defining the solubility limit in this compound series to be between $x = 0.125$ and $x = 0.133$. Diffraction analysis of the $[Bi_{1-x}Mn_x]Te$ series of compounds, which were synthesized with no anti-site defect chemical compensation, also found them to have the BiTe structure (Fig. 3) but to have a much smaller Mn solubility range (up to $x = 0.067$). For larger x, a MnTe impurity forms, defining the solubility limit in that case.

**Magnetic and electronic properties**

*(a) $[Bi_{1-2x}Te_xMn_x]Te$*: The inverse FC susceptibility for x = 0.125 is given in the main panel of Fig. 4. The insets show the low temperature ZFC and FC susceptibility for x = 0.125 and the inverse FC susceptibilities, $1/(\chi(T)-\chi_0)$ for $0.033 \leq x \leq 0.125$. The temperature independent contribution ($\chi_0$, given in Fig. 4), due to core diamagnetism and temperature independent paramagnetic contributions, was obtained by fitting the susceptibility to $\chi(T) = \chi_0 + C/(T-\theta)$, where C is the Curie constant and $\theta$ the Weiss temperature. For x = 0.125, a Curie-Weiss fit in the 30 K $\leq$ T $\leq$ 300 K interval (solid line) gives $\mu_{eff}$ = 4.90 $\mu_B$/Mn and $\theta$ = 8.6 K and $\chi_0$ = 0. The expected (spin-only) values for high-spin $Mn^{3+}$ (S = 2) and $Mn^{2+}$ (S = 5/2) are 4.9 $\mu_B$ and 5.9 $\mu_B$, respectively. From the inset it can be seen that the Weiss temperature is close to zero for x $\leq$ 0.067, while for x > 0.067 positive values are found, signaling the presence of ferromagnetic interactions. The occurrence of a ferromagnetic state for x = 0.100 and x = 0.125 is confirmed by Arrott plots (Fig. 5a,b). Theory predicts that the (high-field) isotherms of $M^2$ vs. H/M are parallel lines for ferromagnets, and that the isotherm at $T_c$ passes through zero.[20] The Curie temperature is close to 10 K for x = 0.125 (Fig. 5a) and close to 5 K for x = 0.100 (Fig. 5b) as the intercepts of the linear fits to the $M^2$ vs. H/M plots at those temperatures are closest to (0,0). The insets in Fig. 5a,b show the as measured M(H) isotherms. Both compositions show magnetic hysteresis at 2 K (Fig. 5c), characteristic of ferromagnetism. For x = 0.125, the coercive field $H_c$ = 330 Oe and the remnant magnetization $M_r$ = 1.28 $\mu_B$/Mn. For x = 0.100, $H_c$ = 79 Oe, and $M_r$ = 0.21 $\mu_B$/Mn. For x = 0.033 and x = 0.067 no hysteresis is observed. The linear increase of the magnetization in high-fields (temperatures below $T_c$, insets in Fig. 5) can be attributed to magneto-crystalline anisotropy: whereas the easy-axis magnetization can saturate within accessible low field ranges, the magnetization along the hard-axes usually increases linearly with field over the same low field range.[20] This is consistent with the finite slope of the field dependent average magnetization as observed in our polycrystalline samples.

*(b) [Bi$_{1-x}$Mn$_x$]Te*: The temperature dependence of the inverse ZFC susceptibilities, $1/(\chi(T)-\chi_0)$ are shown in Fig. 6, and the field dependencies of the magnetization at 5 K are shown in the inset. The inverse susceptibilities almost overlap. Curie-Weiss fitting ($5 \leq T \leq 150$ K) resulted in effective moments of 4.9 $\mu_B$/Mn for both x, while the Weiss temperatures are small and negative ($\theta = -2(1)$ K). The effective moments are close to that observed for [Bi$_{0.75}$Te$_{0.125}$Mn$_{0.125}$]Te. The observed magnetizations at 5 K and 50 kOe are approximately 2.5 $\mu_B$/Mn for both compositions.

A comparison of the effective moments, Weiss temperatures and magnetization at 5 K and 50 kOe for [Bi$_{1-2x}$Te$_x$Mn$_x$]Te and [Bi$_{1-x}$Mn$_x$]Te is given in Fig. 7. The effective moment for [Bi$_{1-2x}$Te$_x$Mn$_x$]Te increases to what appears to be an asymptotic value of 4.9 $\mu_B$/Mn. The same value is found for both [Bi$_{1-x}$Mn$_x$]Te compositions. The Weiss temperature is close to 0 K for $x \leq 0.067$ in both series, while for $x > 0.067$, $\theta$ is positive for [Bi$_{1-2x}$Te$_x$Mn$_x$]Te with a maximum of +8.6 K (x = 0.125). The magnetization at 5 K and 50 kOe increases linearly with x and reaches a maximum of 3.2 $\mu_B$/Mn for x = 0.125. This is smaller than expected for high-spin Mn$^{3+}$ (4 $\mu_B$) and Mn$^{2+}$ (5 $\mu_B$). The present experimental data do not allow for an unambiguous determination of the Mn valence state. The effective moment and magnetization (5 K, 50 kOe) are closer to the expected values for Mn$^{3+}$ than for Mn$^{2+}$. Specifically the effective moment (4.9 $\mu_B$/Mn) is in very good agreement. However, it must be noted that recent results on thin films of Ga$_{1-x}$Mn$_x$As show a magnetization around 4 $\mu_B$/Mn - 4.5 $\mu_B$/Mn after correction for interstitial Mn defects.[2] In the current case, the observed magnetization could also be reduced from its expected value due to compensation from some fraction of the Mn, which as in the case of Ga$_{1-x}$Mn$_x$As may be antiferromagnetically coupled. Investigations into the importance of compensation by interstitial Mn and explanation of the doping dependence of the magnetic properties are for future studies. The main conclusion from this work is the observation of ferromagnetism in [Bi$_{1-2x}$Te$_x$Mn$_x$]Te.

The temperature dependence of the resistivity of [Bi$_{0.75}$Te$_{0.125}$Mn$_{0.125}$]Te is given in Fig. 8 and is typical of a degenerate semiconductor or poor metal. The weak temperature dependence (the residual resistivity ratio is 1.52) indicates charge carrier scattering is dominated by structural defects or impurities, consistent with the random doping of Mn in the (2) blocks and the presence of Te$_{Bi}$ anti-site defects in the (5) blocks. The insets show the temperature dependencies of the Seebeck coefficient (S) and the thermal conductivity (κ). The negative sign of S indicates that n-type conduction is dominant, and the magnitude is comparable to that observed for BiTe. (S$_{RT}$ = -30 µV/K, on our samples). The change of curvature around 20 K coincides with the maximum in thermal conductivity and the minimum in resistivity.

The field dependence of the Hall resistivity of [Bi$_{0.75}$Te$_{0.125}$Mn$_{0.125}$]Te is given in Fig. 9. The data reveals the presence of an anomalous Hall effect, which exists to temperatures above T$_c$. In metals, ferromagnetism is invariably accompanied by the existence of a large anomalous Hall effect. The observed Hall resistivity ρ$_{xy}$ is comprised of two contributions, viz.

$$\rho_{xy} = R_0 H + R_s M, \qquad (1)$$

where R$_0$ and R$_s$ are the ordinary and anomalous Hall coefficients, respectively, and M is the magnetization. The strongly nonlinear field profile of ρ$_{xy}$ is made up of the linear term R$_0$H and the nonlinear term R$_s$M that reflects the H dependence of M. In the present system, the anomalous term is strongly in evidence at 4.5 K, and is superposed on an *n*-type H linear term. We have found that, above T$_c$, the anomalous term remains quite sizable (e.g. see the 15K data in Fig. 9). A detailed study of this will be reported elsewhere.[21] The interesting extension of the anomalous term high above T$_c$ is also observed in other ferromagnetic systems including $Ga_{1-x}Mn_xAs$.[22] By separating the ordinary term $R_0$ in Eq. 1, we have derived the carrier density $n$ in $[Bi_{0.75}Te_{0.125}Mn_{0.125}]Te$. Figure 9 shows that, over a broad interval of T, $n$ has nearly constant value $\sim 7 \times 10^{20}$ cm$^{-3}$. This is two orders of magnitude higher than the value reported in $Bi_2Te_3$.[23]

**Conclusions**

The substitution of Mn in BiTe is demonstrated for the first time. Structural analysis for $[Bi_{0.75}Te_{0.125}Mn_{0.125}]Te$ shows that Mn preferentially occupies sites in the $Bi_2$ blocks and is accompanied by an equal amount of $Te_{Bi}$ anti-site defects. These anti-site defects greatly increase the solubility of Mn, as demonstrated by the relative phase stabilities of the $[Bi_{1-x}Mn_x]Te$ and $[Bi_{1-2x}Te_xMn_x]Te$ series. This indicates that electronic considerations significantly influence the crystal chemistry of these systems. $[Bi_{1-x}Mn_x]Te$ shows little change in magnetic properties with x and remains paramagnetic down to 5 K. The magnetism in $[Bi_{1-2x}Te_xMn_x]Te$ on the other hand shows a clear doping dependence, and, for x larger than 0.067 ferromagnetism is observed. From simple chemical arguments (Te has one more valence electron than Bi) the $Te_{Bi}$ anti-site defects in the $Bi_2Te_3$ blocks are expected to be single donors. The effect of Mn substitution in the $Bi_2$ layers is harder to predict but it is likely that Mn acts as an acceptor and the $Te_{Bi}$ anti-site defects and $Mn_{Bi}$ partially compensate each other, resulting in a greater Mn solubility.

$[Bi_{0.75}Te_{0.125}Mn_{0.125}]Te$ has been studied in most detail and has a $T_c$ close to 10 K. In this case, the composition of the $Bi_2$ block is approximately $(Mn_{0.4}Bi_{0.6})_2$. Such large Mn concentrations in neutral hosts does not automatically result in ferromagnetism; as demonstrated by $Zn_{1-x}Mn_xTe$ and $Cu_{1-x}Mn_x$, which are spin-glasses. On the other hand, ferromagnetism with $T_c$ up to 170 K is observed in $Ga_{1-x}Mn_xAs$. In the latter case, the introduction of Mn is accompanied by charge carrier doping and this appears vital for the occurrence of ferromagnetism. The localization of Mn in the two-layer wide Bi blocks in BiTe, creating layers of relatively high Mn concentration, and the charge compensation due to the Te$_{Bi}$ anti-site defects, appear to be important factors in why Mn-doped BiTe becomes ferromagnetic at low temperatures. The fact that Mn-doped BiTe is a unique bulk system in which structural analysis is possible, showing the locations of both the doped Mn and charge compensating defects, suggest that it is worthy of further study as a model for transition metal doping induced ferromagnetism in semiconductors provided single crystals can be grown.

## Acknowledgements

This research was supported in part by the MRSEC program of the National Science Foundation, and in part by the Air Force Research Laboratory.

Table I. Refined atomic coordinates, temperature factors and occupancies for Mn doped BiTe. The refined composition corresponds to [$Mn_{0.128(3)}Bi_{0.75}Te_{0.122(3)}$]Te.

| Atom | Pos. | x | y | z | $U_{iso}$ (Å$^2$) | Occupancy |
|---|---|---|---|---|---|---|
| (Bi/Mn)1 | 2d | 2/3 | 1/3 | 0.4620(2) | 0.0130(6) | 0.62(1)/0.38(1) |
| (Bi/Te)2 | 2d | 1/3 | 2/3 | 0.2890(1) | 0.0130(6) | 0.63(1)/0.37(1) |
| Bi3 | 2c | 0 | 0 | 0.1143(1) | 0.0130(6) | 1.00 |
| Te1 | 2c | 0 | 0 | 0.3759(2) | 0.0130(6) | 1.00 |
| Te2 | 2d | 2/3 | 1/3 | 0.1898(2) | 0.0130(6) | 1.00 |
| Te3 | 2d | 1/3 | 2/3 | 0.0454(2) | 0.0130(6) | 1.00 |

Space group P-3m1: $a$ = 4.3783(1) Å, $c$ = 23.8107(8) Å. Residuals for the fit: $\chi^2$ = 2.1, $wR_p$ = 11.4% $R_p$ = 8.8%, $R_F^2$ = 9.0%. (Lattice constants for BiTe: $a$ = 4.423(2) Å, $c$ = 24.002(6) Å)[24]

Table II. Selected bond lengths (Å) for [$Mn_{0.128(3)}Bi_{0.75}Te_{0.122(3)}$]Te and corresponding bond lengths for BiTe (from Ref 24).

|  | Mn-BiTe | BiTe |
|---|---|---|
| (Bi/Mn)1-(Bi/Mn)1 | 3.107(4) | 3.267(6) |
| (Bi/Mn)1-Te1 | 3.255(5) | 3.326(6) |
| Te1-(Bi/Te)2 | 3.268(4) | 3.166(9) |
| (Bi/Te)2-Te2 | 3.459(4) | 3.137(5) |
| Te2-Bi3 | 3.102(3) | 3.355(6) |
| Bi3-Te3 | 3.013(4) | 3.044(5) |
| Te3-Te3 | 3.327(7) | 3.679(6) |

**Figure Captions**

Fig. 1. Observed (crosses), calculated (solid line) and difference PXD Rietveld profiles for [$Bi_{0.75}Te_{0.125}Mn_{0.125}$]Te. Reflections markers correspond to the Bragg positions.

Fig. 2. View of the structure of [$Bi_{0.75}Te_{0.125}Mn_{0.125}$]Te. The labeling corresponds to that used in Tables 1 and 2 and throughout the text.

Fig. 3. PXD patterns for [$Bi_{1-2x}Te_xMn_x$]Te and [$Bi_{1-x}Mn_x$]Te.

Fig. 4. Magnetic susceptibilities for [$Bi_{1-2x}Te_xMn_x$]Te. The main panel shows the temperature dependence of the inverse FC susceptibility for $x = 0.125$. The insets show the low temperature ZFC and FC susceptibility for $x = 0.125$ and inverse low temperature FC susceptibilities for $0.033 \leq x \leq 0.125$.

Fig. 5a, b. Arrot plots for [$Bi_{0.75}Te_{0.125}Mn_{0.125}$]Te and [$Bi_{0.80}Te_{0.100}Mn_{0.100}$]Te, respectively. The insets show the as measured M(H) isotherms. (c) Magnetic hysteresis curves for [$Bi_{1-2x}Te_xMn_x$]Te at 2 K. The inset shows the M(H) dependence up to 50 kOe.

Fig. 6. Temperature dependence of the inverse susceptibility for [$Bi_{1-x}Mn_x$]Te. The inset shows the M(H) dependence up to 50 kOe.

Fig. 7. Comparison of the effective moments, Weiss temperatures and magnetization at 5 K and 50 kOe for [$Bi_{1-2x}Te_xMn_x$]Te and [$Bi_{1-x}Mn_x$]Te.

Fig. 8. Temperature dependence of the electrical resistivity for [Bi$_{0.75}$Te$_{0.125}$Mn$_{0.125}$]Te. The inset shows the temperature dependencies of the thermal conductivity (top) and thermopower (bottom).

Fig. 9. Field dependence of the Hall resistivity at selected temperatures. The inset shows the temperature dependence of the carrier density determined from the ordinary Hall coefficient.

Fig. 1

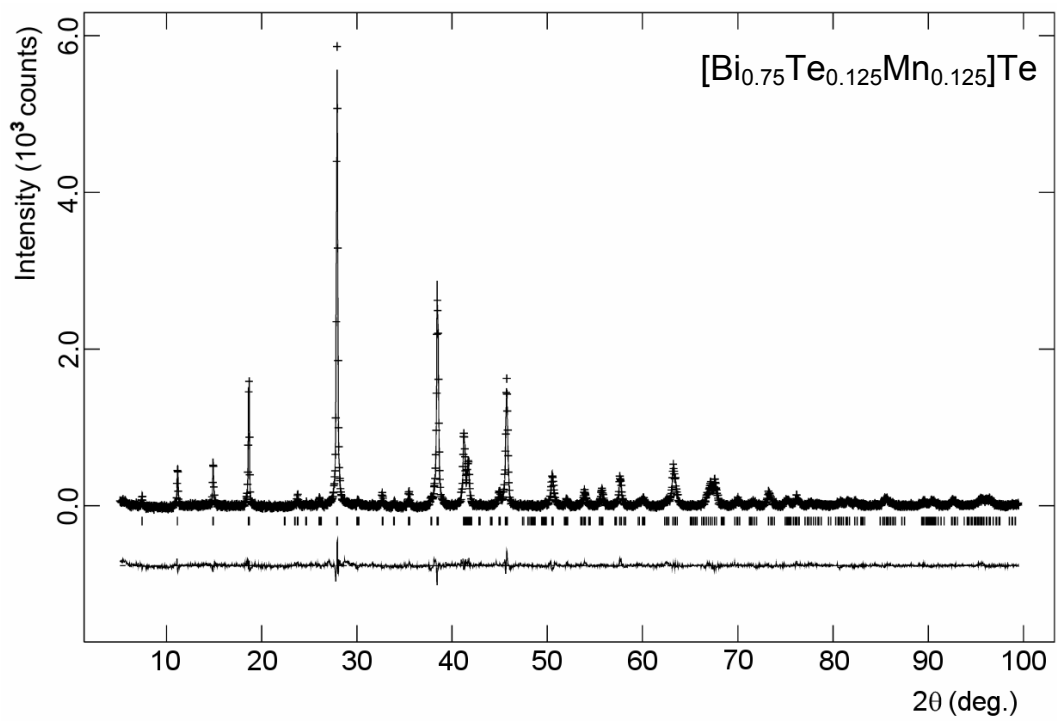

Fig. 2.

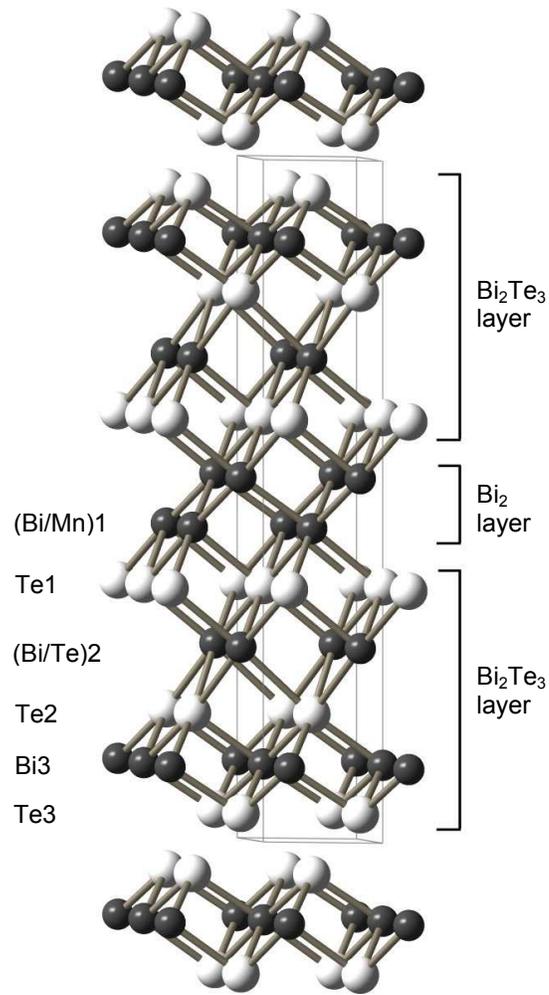

Fig. 3.

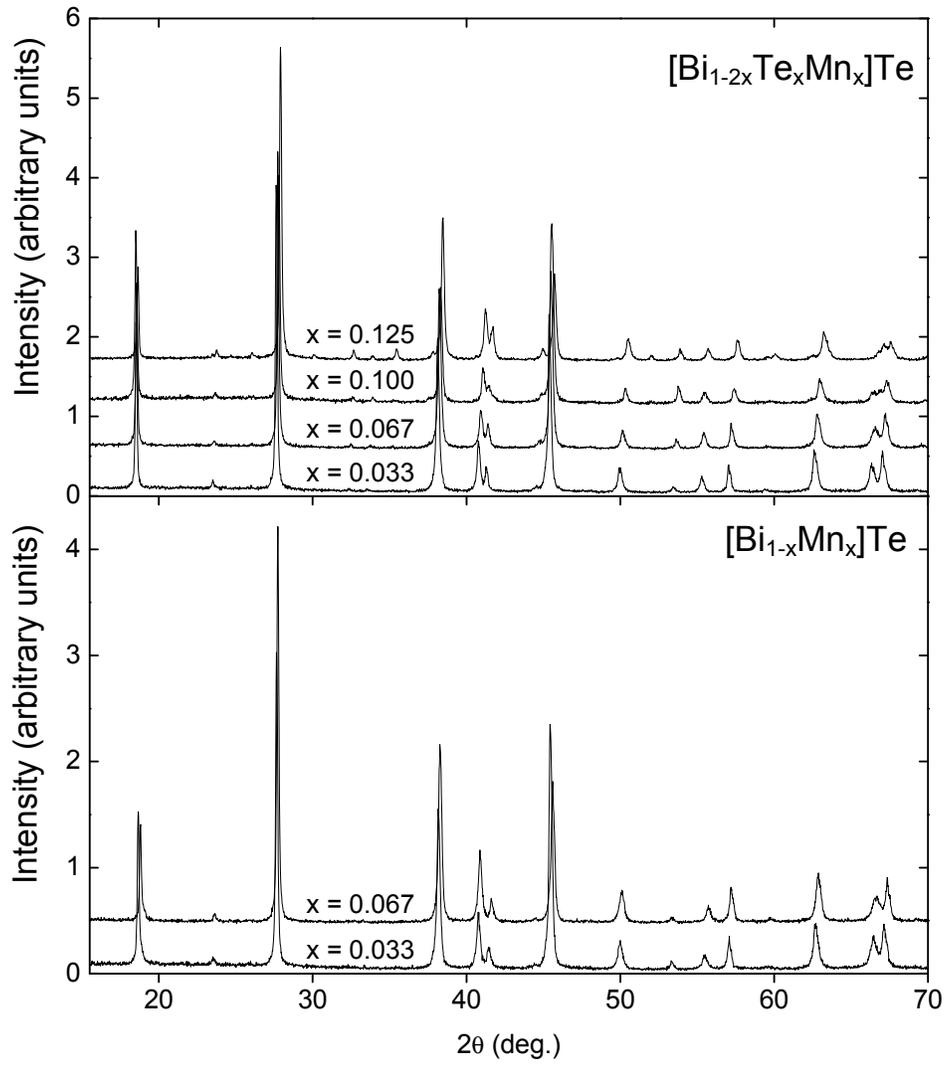

Fig. 4

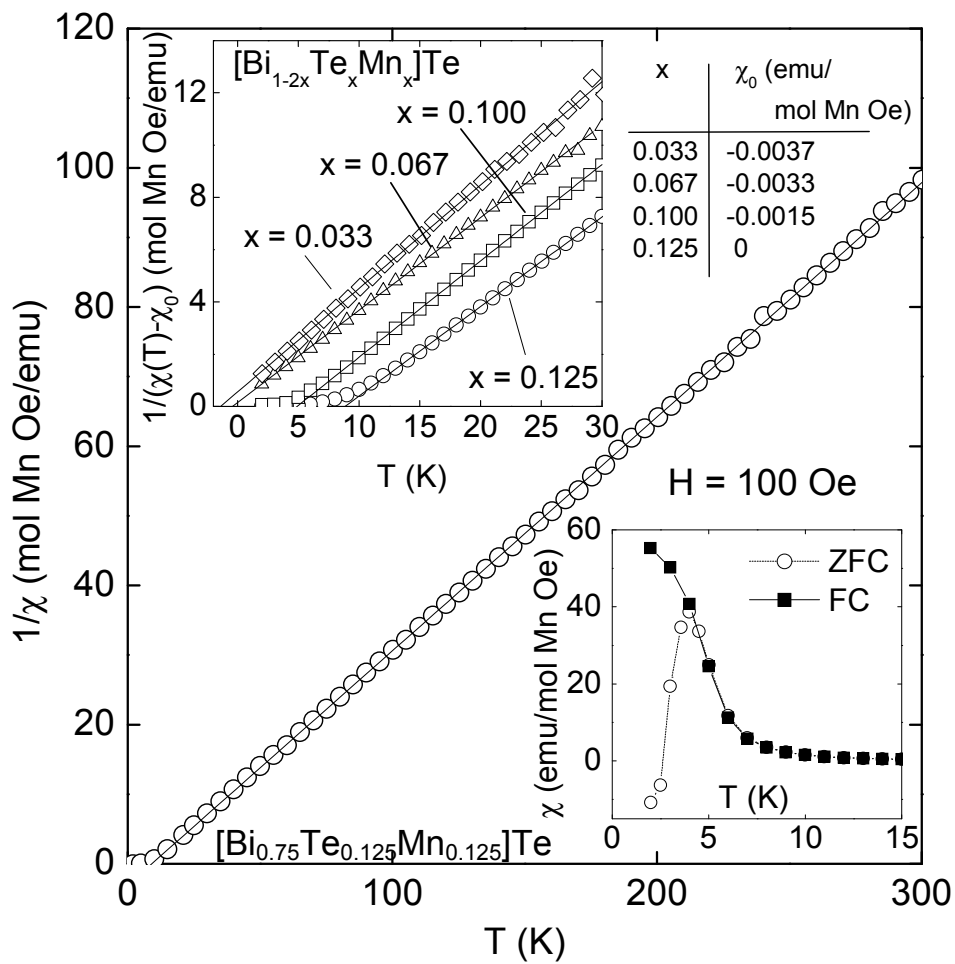

Fig. 5a

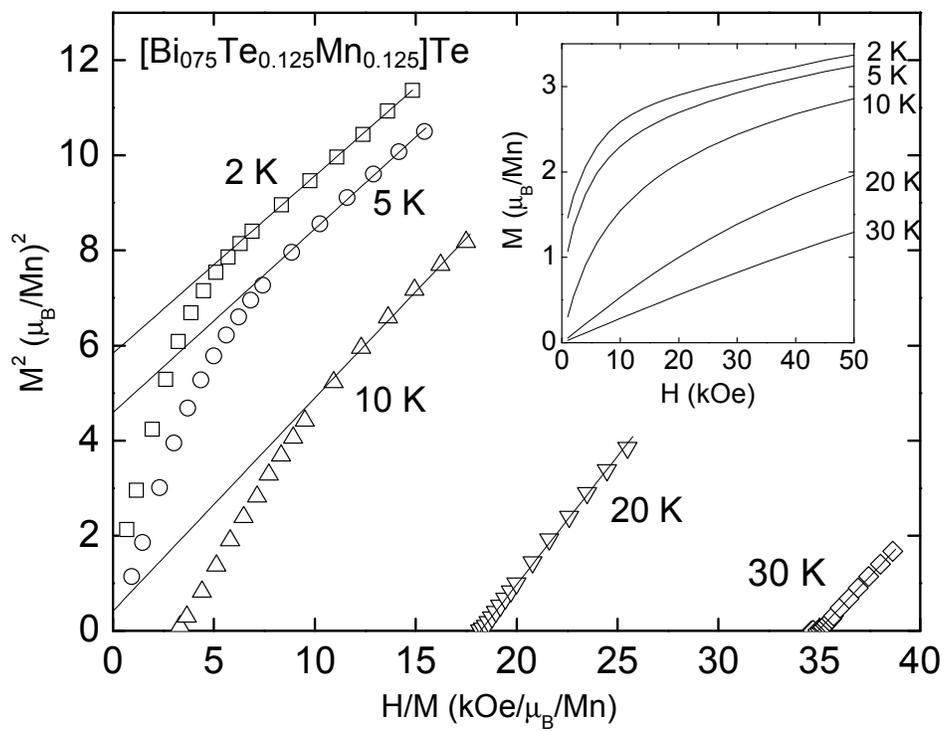

Fig. 5b

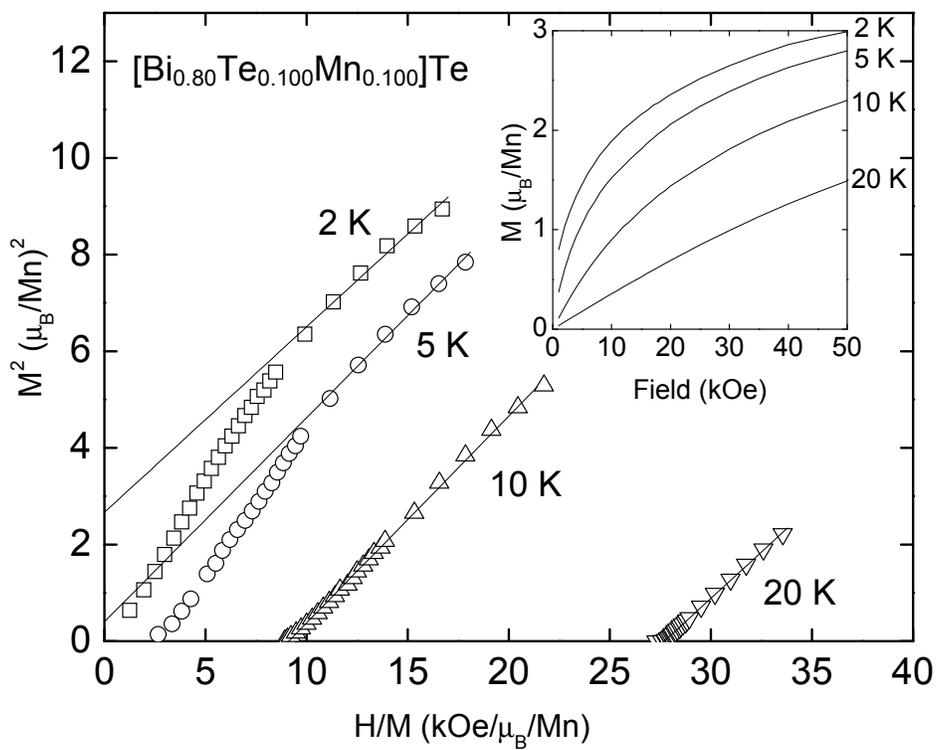

Fig. 5c

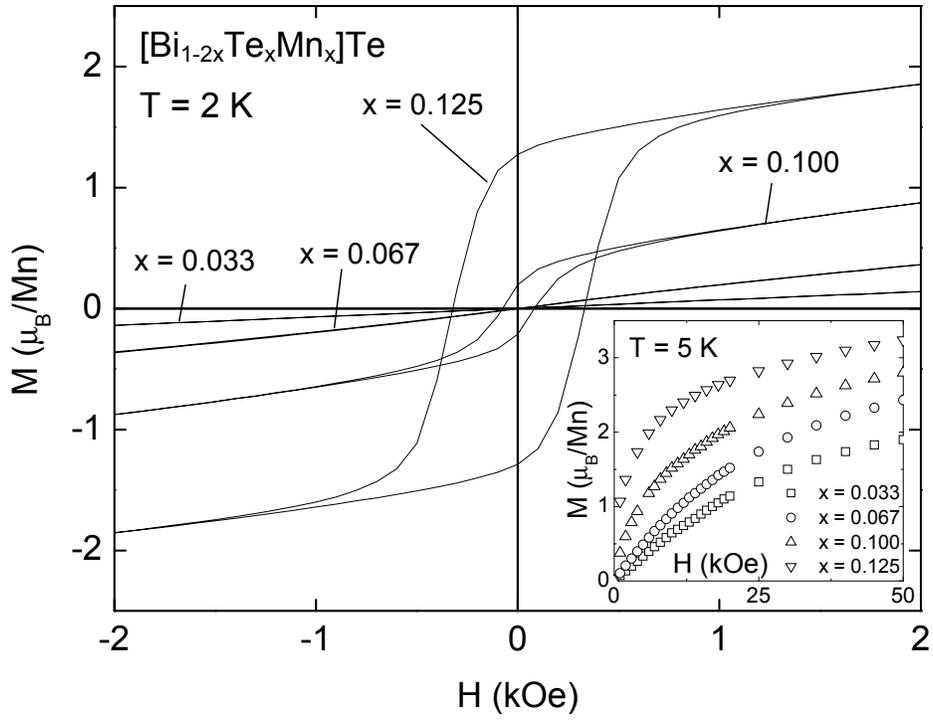

Fig. 6

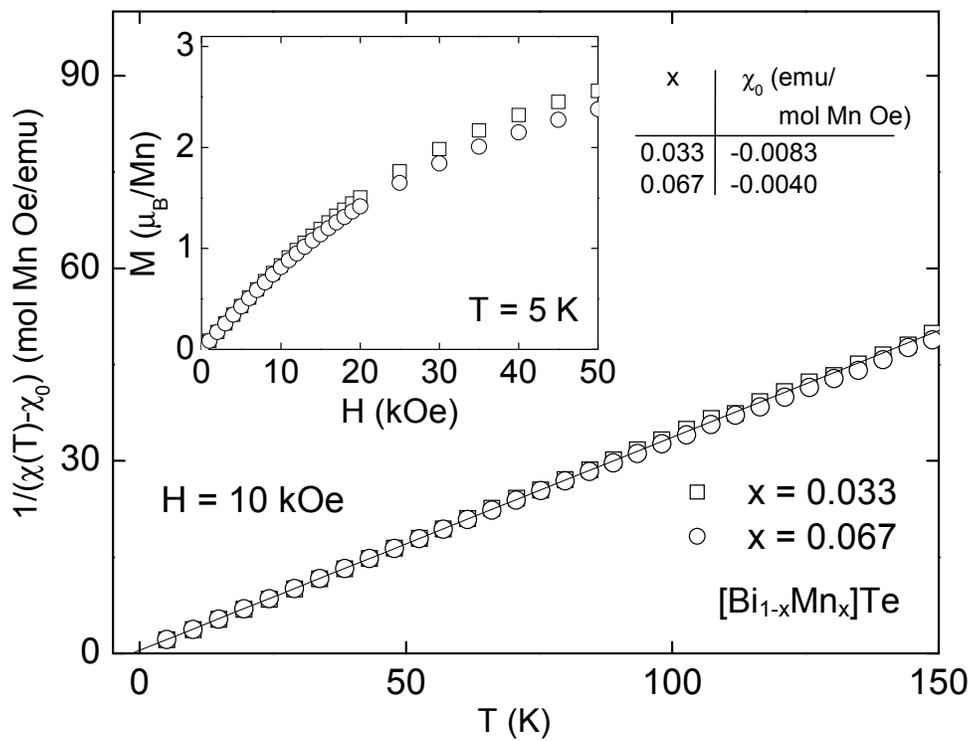

Fig. 7.

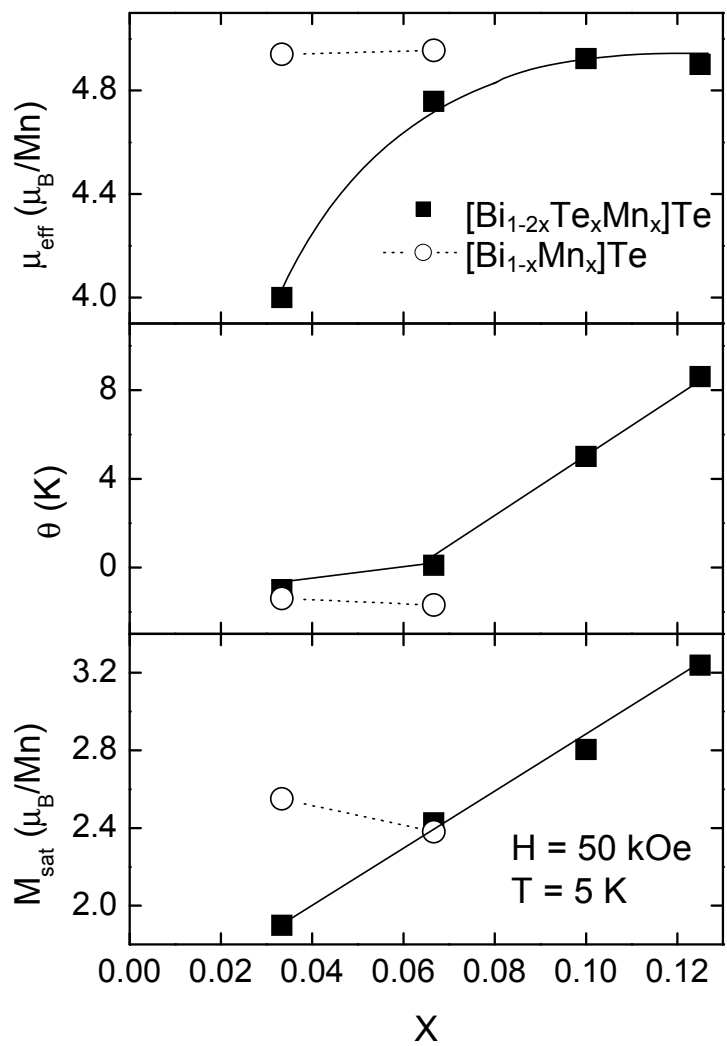

Fig. 8.

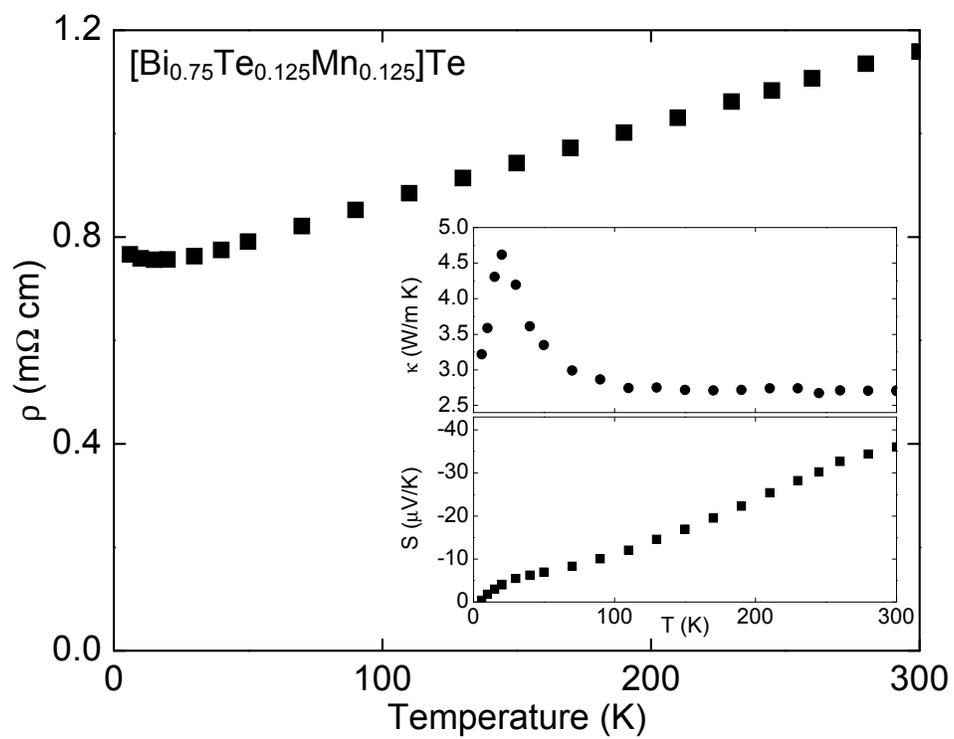

Fig. 9

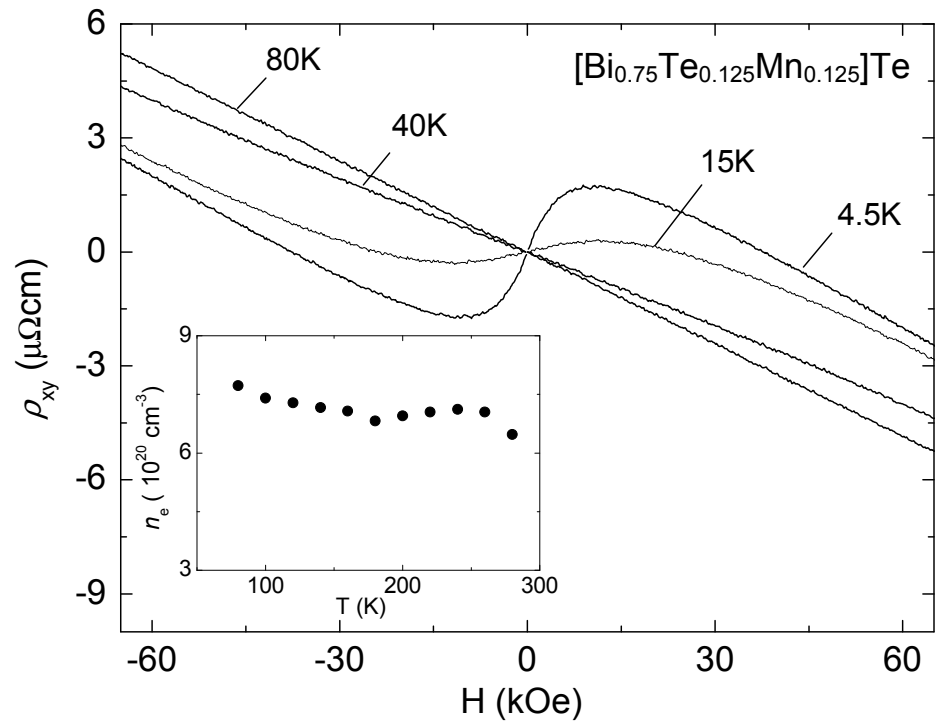